# Hierarchical Temperature Imaging Using Pseudo-Inversed Convolutional Neural Network Aided TDLAS Tomography

Jingjing Si, *Member, IEEE*, Guoliang Li, Yinbo Cheng, Rui Zhang, Godwin Enemali, *Member, IEEE*, Chang Liu, *Member, IEEE*

*Abstract*—As an *in situ* combustion diagnostic tool, Tunable Diode Laser Absorption Spectroscopy (TDLAS) tomography has been widely used for imaging of two-dimensional temperature distributions in reactive flows. Compared with the computational tomographic algorithms, Convolutional Neural Networks (CNNs) have been proofed to be more robust and accurate for image reconstruction, particularly in case of limited access of laser beams in the Region of Interest (RoI). In practice, flame in the RoI that requires to be reconstructed with good spatial resolution is commonly surrounded by low-temperature background. Although the background is not of high interest, spectroscopic absorption still exists due to heat dissipation and gas convection. Therefore, we propose a Pseudo-Inversed CNN (PI-CNN) for hierarchical temperature imaging that (a) uses efficiently the training and learning resources for temperature imaging in the RoI with good spatial resolution, and (b) reconstructs the less spatially resolved background temperature by adequately addressing the integrity of the spectroscopic absorption model. In comparison with the traditional CNN, the newly introduced pseudo inversion of the RoI sensitivity matrix is more penetrating for revealing the inherent correlation between the projection data and the RoI to be reconstructed, thus prioritising the temperature imaging in the RoI with high accuracy and high computational efficiency. In this paper, the proposed algorithm was validated by both numerical simulation and lab-scale experiment, indicating good agreement between the phantoms and the high-fidelity reconstructions.

## I. INTRODUCTION

IN the past decade, Tunable Diode Laser Absorption Spectroscopy (TDLAS) tomography has been developed as a robust optical modality for non-invasive imaging of the cross-sectional distributions of critical parameters, e.g. temperature [1], species concentration [2, 3], and velocity [4], in reactive flows. The mathematical formulation of TDLAS tomography is similar as hard-field computer tomography, but commonly suffers from a small number of projections due to the limited access of laser beams in the sensing region. As a result, the inverse problem of TDLAS tomography is inherently ill-posed with severe rank deficiency [5].

The computational tomographic algorithms, e.g. algebraic techniques [6], entropy regularization [7], statistical inversion [8] and global optimization [9], can mitigate the ill-posedness by imposing *a priori* smoothness of the target field. However, these algorithms inevitably suffer from complex and time-consuming computation, and cannot completely suppress the artefacts where few spatial sampling exists. In contrast, the data-driven algorithms [10-12], empowered by hardware acceleration [13], can significantly speed up the tomographic image reconstruction. With appropriate establishment of training sets, the data-driven algorithms contribute to high accuracy of the image reconstruction, and, in particular, strong robustness to the noise contaminated in tomographic TDLAS measurements.

As one of the most representative data-driven algorithms, Convolutional Neural Networks (CNNs) have been recently validated to solve the inverse problems of TDLAS tomography. Pioneering works mainly considered the cases of densely spatial sampling [14, 15], that is, beam arrangements with at least 6 angular projection views and tens of laser beams per view. To address the practical deployment of TDLAS tomographic sensor with limited optical access, we recently demonstrated a novel CNN-aided algorithm to reconstruct the distributions of flame temperature and species concentration with only 32 laser beams, that is, 4 angular projection views and 8 laser beams per view [16]. Among all the above efforts, the sensing region was segmented by uniform-size pixels, resulting into uniformly allocated training and learning resources for each pixel. However, many industrial combustors are bypassed by cooling gas flow, for example, the gas turbine exhaust [17]. The combustion zone, known as the Region of Interest (RoI), needs to be finely resolved, while the spectroscopic absorption in the cooling background should also be considered but is of

This work was supported in part by the National Natural Science Foundation of China under Grant 61701429, and the U.K. Engineering and Physical Sciences Research Council under Platform Grant EP/P001661/1. (*Corresponding author: Chang Liu.*)

J. Si and G. Li are with the School of Information Engineering, Yanshan University, Qinhuangdao 066004, China.

Y. Cheng is with the Ocean College, Hebei Agricultural University, Qinhuangdao 066003, China.

R. Zhang, G. Enemali, and C. Liu are with the School of Engineering, University of Edinburgh, Edinburgh EH9 3JL, U.K. (e-mail: C.Liu@ed.ac.uk).

less interest. In these scenarios, the requirements placed on CNN are to (a) allocate the majority of the computational force in the RoI and (b) adequately consider the background absorption with minor hardware resources.

To address the above-mentioned issues, we propose a hierarchical imaging strategy using Pseudo-Inversed CNN (PI-CNN) aided TDLAS tomography in this paper. Dense meshes are introduced in the RoI to prioritize detailed reconstruction of the target flow, while spares ones are used in the background to address the integrity of the physical model. Instead of allocating the computational force uniformly in the entire sensing region, the computational force is used more efficiently in the proposed method with a stronger weight in the RoI over the background. Compared with the traditional CNN, the introduced pseudo inversion is more competent at revealing the inherent correlation between the projection data and the true image with hierarchical information.

The remainder of this paper is organized as follows. Based on the mathematical formulation of TDLAS tomography, we firstly introduce the hierarchical imaging strategy in Section II. Then, the PI-CNN architectures is established using the 32-beam TDLAS tomographic sensor in Section III. Subsequently, the established network is trained with its performance examined by simulated test set and experiments in Sections IV and V, respectively. Finally, a brief conclusion is presented in Section VI.

## II. Hierarchical Imaging Strategy

The mathematical formulation of TDLAS tomography is firstly reviewed in this section to facilitate the introduction of the hierarchical imaging strategy. When a laser beam at frequency $v$ [cm$^{-1}$] permeates an inhomogeneous absorbing medium with a path of length $L$ [cm], the path integrated absorbance $A_v$ is described as:

$$A_v = \int_0^L a_v(l)dl = \int_0^L [P(l)\cdot S_v(T(l))\cdot X(l)]dl, \quad (1)$$

where $P(l)$ [atm], $T(l)$ [K] and $X(l)$ are the local pressure, temperature and molar fraction of the absorbing species (gas concentration), respectively. $S_v(\cdot)$ [cm$^{-2}$atm$^{-1}$] denotes the temperature-dependent line strength [18]. $a_v(l)$ is defined as local absorbance density that satisfies $a_v(l)=P(l)\cdot S_v(T(l))\cdot X(l)$.

The inverse problem of TDLAS tomography is formulated by discretizing the sensing region into $J$ pixels with uniform gas parameters assumed in each pixel [19], as shown in Fig. 1. The sensing region is defined as octagonal in Fig. 1, which is consistent with the optical layout introduced later in Section III. As a result, the discrete expression of (1) can be written as

$$A_v = L a_v, \quad (2)$$

where $A_v \in \mathbb{R}^{I\times 1}$ denotes the path integrated absorbances obtained from a total of $I$ Line-of-Sight (LoS) TDLAS measurements, with its element $A_{v,i}$ representing the path integrated absorbance of the $i$-th beam. $L \in \mathbb{R}^{I\times J}$ is the sensitivity matrix with its element $L_{i,j}$ representing the chord length of the laser path for the $i$-th laser beam passing through the $j$-th pixel. $i\in\{1, 2,…, I\}$ and $j\in\{1, 2,…, J\}$ are the indices of laser beams and pixels, respectively. $a_v \in \mathbb{R}^{J\times 1}$ is the vector of absorbance density with its elements $a_{v,j}$ expressed as

$$a_{v,j} = P_j X_j S_v(T_j). \quad (3)$$

To resolve detailed flow parameters in the combustion field, in this work, a total of $J^{RoI}$ dense meshes are introduced in the RoI. In contrast, the less interested background is coarsely segmented, resulting into $J-J^{RoI}$ pixels. Here, $J^{RoI}$ is much larger than $J-J^{RoI}$ to give the RoI higher hierarchy for using computational force. The sensitivity matrix $L$ can be described as

$$L = \begin{bmatrix} L_{1,1}^{RoI} & \cdots & L_{1,J^{RoI}}^{RoI} & L_{1,1}^{BG} & \cdots & L_{1,J-J^{RoI}}^{BG} \\ \vdots & \ddots & \vdots & \vdots & \ddots & \vdots \\ L_{I,1}^{RoI} & \cdots & L_{I,J^{RoI}}^{RoI} & L_{I,1}^{BG} & \cdots & L_{I,J-J^{RoI}}^{BG} \end{bmatrix}, \quad (4)$$

$$\underbrace{\quad\quad\quad\quad}_{L^{RoI}\in\mathbb{R}^{I\times J^{RoI}}} \underbrace{\quad\quad\quad\quad}_{L^{BG}\in\mathbb{R}^{I\times (J-J^{RoI})}}$$

where $L_{i,j}^{RoI}$ and $L_{i,j}^{BG}$ are the elements in $L^{RoI}$ and $L^{BG}$, denoting the chord lengths of the $i$-th laser path in the $j$-th pixel in the RoI and the background, respectively. Similarly, the vector of hierarchical absorbance density $a_v$ can be expressed as

$$a_v = [\underbrace{a_{v,1}^{RoI} \cdots a_{v,J^{RoI}}^{RoI}}_{a_v^{RoI}\in\mathbb{R}^{J^{RoI}\times 1}} \underbrace{a_{v,1}^{BG} \cdots a_{v,J-J^{RoI}}^{BG}}_{a_v^{BG}\in\mathbb{R}^{(J-J^{RoI})\times 1}}]^T, \quad (5)$$

where $a_{v,j}^{RoI}$ and $a_{v,j}^{BG}$ are the elements in $a_v^{RoI}$ and $a_v^{BG}$, denoting the absorbance densities in the $j$-th pixel in the RoI and the background, respectively.

By substituting (3) into (5), $a_v$ can be expressed as

$$a_v = \begin{bmatrix} a_v^{RoI} \\ a_v^{BG} \end{bmatrix} = P[S_v(T)\odot X] = P\begin{bmatrix} S_v(T^{RoI})\odot X^{RoI} \\ S_v(T^{BG})\odot X^{BG} \end{bmatrix}, \quad (6)$$

where $T=\begin{bmatrix} T^{RoI} \\ T^{BG} \end{bmatrix}$ and $X=\begin{bmatrix} X^{RoI} \\ X^{BG} \end{bmatrix}$ are the vectors of hierarchical temperature and species concentration, respectively. $T^{RoI}\in\mathbb{R}^{J^{RoI}\times 1}$ ($X^{RoI}\in\mathbb{R}^{J^{RoI}\times 1}$) and $T^{BG}\in\mathbb{R}^{(J-J^{RoI})\times 1}$ ($X^{BG}\in\mathbb{R}^{(J-J^{RoI})\times 1}$) are the sub-vectors corresponding to the RoI and the background, respectively. $P$ is the local pressure assumed uniform in the sensing region. $\odot$ denotes element-wise multiplication.

By substituting (4) and (6) into (2), the hierarchical imaging problem can be finally expressed as

$$A_v = [L^{RoI}\ L^{BG}]\begin{bmatrix} a_v^{RoI} \\ a_v^{BG} \end{bmatrix} = L^{RoI}a_v^{RoI} + L^{BG}a_v^{BG}$$
$$= PL^{RoI}[S_v(T^{RoI})\odot X^{RoI}] + PL^{BG}[S_v(T^{BG})\odot X^{BG}]. \quad (7)$$

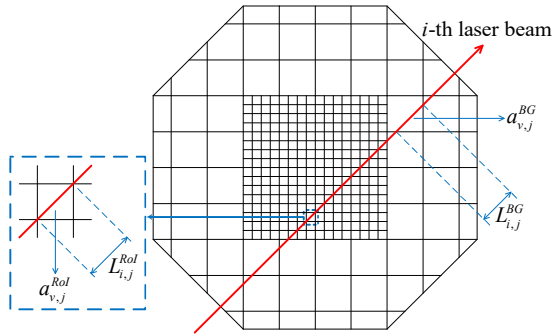

Fig. 1. Geometric description of a TDLAS line-of-sight measurement in an octagonal sensing region with hybrid-size pixels.

To reconstruct $T^{RoI}$ ($X^{RoI}$) and $T^{BG}$ ($X^{BG}$), two absorbing transitions for the same species will be adopted to cover the target temperature range with adequate temperature sensitivity, i.e. the so called two-line thermometry [19]. In this work, the hierarchical temperature distribution $T=\begin{bmatrix} T^{RoI} \\ T^{BG} \end{bmatrix}$ is reconstructed, as an example, based on the PI-CNN proposed in Section III, from path integrated absorbances measured at the two selected absorbing transitions.

### III. PSEUDO-INVERSED CNN ARCHITECTURES

To facilitate the presentation of the proposed PI-CNN, the optical layout of our lab-scale 32-beam TDLAS tomographic senor is firstly introduced in this section. As shown in Fig. 2, the 32 laser beams are arranged in 4 equiangular projection angles at 0°, 45°, 90° and 135°, each angle with 8 equispaced parallel beams. The distance between adjacent beams is 18 mm, which makes the octagonal sensing region with side length of 144 mm. As a major product of combustion, water-vapor ($H_2O$) is chosen as the target absorbing species for imaging of the flow parameters. Two absorbing transitions at $v_1$=7185.6 cm$^{-1}$ and $v_2$=7444.36 cm$^{-1}$ are selected due to their moderate line strengths and good temperature sensitivity over the target temperature range of 300-1200 K [20]. More details of the optics and electronics were described in our previous publications [7, 21].

In this work, the central square area of the sensing region with dimensions of 144 mm × 144 mm is chosen as the RoI, while the area out of the RoI is regarded as the background. The RoI is discretized into 40 × 40 pixels ($J^{RoI}$=1600), each with dimensions of 3.6 mm × 3.6 mm. In contrast, the background is coarsely discretized, containing 364 pixels ($J-J^{RoI}$=364), each with dimensions of 14.4 mm × 14.4 mm. The proposed PI-CNN is constructed to reconstruct hierarchical temperature $T=\begin{bmatrix} T^{RoI} \\ T^{BG} \end{bmatrix} \in \mathbb{R}^{1964\times1}$ from path integrated absorbances $A_{v_1} \in \mathbb{R}^{32\times1}$ and $A_{v_2} \in \mathbb{R}^{32\times1}$ measured by the 32-beam sensor at $v_1$ and $v_2$, respectively.

Due to the small number of projection data in $A_{v_1}$ and $A_{v_2}$, the inverse problem here is severely rank deficient, that is, the number of measurements is far less than the number of unknown parameters to be retrieved. When implementing the deep learning algorithm, the rank deficiency imposes significant challenges to perceive distributions of flow parameters, particularly in the RoI, from $A_{v_1}$ and $A_{v_2}$. To further reveal the inherent correlation between TDLAS measurements and the true distributions of flow parameters, pseudo inversion of $L^{RoI}$, noted as $[L^{RoI}]^+$, is introduced in the proposed deep learning algorithm. The pseudo inversion is to obtain a pre-reconstructed image in the RoI with minimal computational force, enabling closer correlation to the true image in comparison with the raw measurements. For example, Fig. 3 (a), with 40×40 pixels, shows the true distribution of the absorbance densities in the RoI $a_{v_1}^{RoI} \in \mathbb{R}^{1600\times1}$ obtained from a single-peak phantom presented later in Section IV, while Fig. 3 (b), with 8×4 pixels demonstrates the normalized 32-beam measurements $A_{v_1} \in \mathbb{R}^{32\times1}$. Apparently, no spatially resolved pattern of $a_{v_1}^{RoI}$ can be recognized from $A_{v_1}$. In contrast, Fig. 3 (c) shows the pre-reconstructed image $C_{v_1} \in \mathbb{R}^{1600\times1}$ generated by $C_{v_1}=[L^{RoI}]^+ A_{v_1}$. Although Fig. 3 (c) suffers from ray-like artefacts, it indicates the rough location of the peak in the phantom that would strongly assist to perceive the true distribution of $a_{v_1}^{RoI}$ shown in Fig. 3 (a).

By integrating the above-mentioned pseudo inversion in CNN, the architecture of the deep-learning model shown in Fig. 4, named as Pseudo-Inversed CNN (PI-CNN), is constructed by the following three steps.

Firstly, the pseudo inversion (PI) layer multiplies $[L^{RoI}]^+$ with the inputs $A_{v_1}$ and $A_{v_2}$, and reshapes the generated $C_{v_1} \in \mathbb{R}^{J^{RoI}\times1}$ and $C_{v_2} \in \mathbb{R}^{J^{RoI}\times1}$, respectively. For each transition $v$, $C_v$ can be expressed as

$$C_v = [L^{RoI}]^+ A_v = [L^{RoI}]^+ L a_v$$
$$= [L^{RoI}]^+ L^{RoI} a_v^{RoI} + [L^{RoI}]^+ L^{BG} a_v^{BG}$$
$$= P[L^{RoI}]^+ L^{RoI} [S_v(T^{RoI}) \odot X^{RoI}] + P[L^{RoI}]^+ L^{BG} [S_v(T^{BG}) \odot X^{BG}]. \tag{8}$$

In (8), the pseudo inversion only involves the chord lengths matrix in the RoI, i.e. $L^{RoI}$, rather than that in the background, i.e. $L^{BG}$. This operation facilitates the CNN to capture the prominent spatial feature in the RoI by coarsely pre-reconstructing $a_v^{RoI}$ with $[L^{RoI}]^+ L^{RoI} a_v^{RoI}$, thus imposing stronger dominance of the RoI in image reconstruction. According to Algorithm 1, $C_{v_1}$ and $C_{v_2}$ are subsequently reshaped into $C_{v_1}^{reshape} \in \mathbb{R}^{H_{RoI} \times W_{RoI}}$ and $C_{v_2}^{reshape} \in \mathbb{R}^{H_{RoI} \times W_{RoI}}$, respectively. Here, $H_{RoI}$ and $W_{RoI}$ denote the height and width of the RoI, respectively.

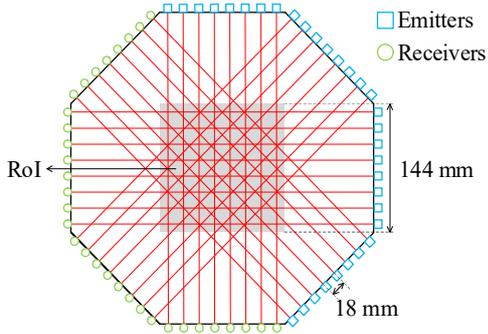

Fig. 2. Beam arrangement for the TDLAS tomographic sensor in the sensing region.

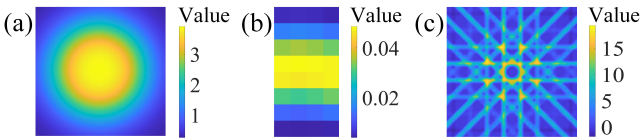

Fig. 3. Exampled pseudo inversion of the sensitivity matrix $L^{RoI}$ with its effect on perceiving the true distributions of flow parameters. (a) shows the true distribution of the absorbance densities $a_{v_1}^{RoI}$ obtained from a single-peak phantom presented later in Section IV; (b) the normalized 32-beam measurements $A_{v_1}$; (c) the pre-reconstructed image $C_{v_1}=[L^{RoI}]^+ A_{v_1}$.

**Algorithm 1** Pre-reconstruction of the RoI in the PI layer

**Input**: Number of TDLAS measurements $I$, dimensions of the RoI $H_{RoI} \times W_{RoI}$, number of pixels in RoI $J^{RoI}$, path integrated absorbance $A_{v_1} \in \mathbb{R}^{I \times 1}$, $A_{v_2} \in \mathbb{R}^{I \times 1}$, and chord lengths matrix $L^{RoI} \in \mathbb{R}^{I \times J^{RoI}}$.

**Operations**:
1: $L^+ \leftarrow$ calculate the pseudo inversion $[L^{RoI}]^+$ of $L^{RoI}$
2: $C_{v_1} \leftarrow L^+ A_{v_1}$
   $C_{v_2} \leftarrow L^+ A_{v_2}$
3: $C_{v_1}^{reshape} \leftarrow$ reshape($C_{v_1}$, $(H_{RoI}, W_{RoI})$)
   $C_{v_2}^{reshape} \leftarrow$ reshape($C_{v_2}$, $(H_{RoI}, W_{RoI})$)

**Output**: $C_{v_1}^{reshape} \in \mathbb{R}^{H_{RoI} \times W_{RoI}}$ and $C_{v_2}^{reshape} \in \mathbb{R}^{H_{RoI} \times W_{RoI}}$

TABLE I
HYPER-PARAMETERS OF THE PROPOSED PI-CNN ARCHITECTURE.

| PI-layer | | | | |
|---|---|---|---|---|
| Number of TDLAS measurements at a certain transition | RoI dim. | RoI chord lengths matrix dim. | Input dim. | Output dim. |
| 32 | 40×40 | 32×1600 | 32×1×2 | 40×40×2 |

| CNN layers | | | | |
|---|---|---|---|---|
| | Input dim. | Output dim. | weight matrix size | Stride | Padding |
| Conv1 | 40×40×2 | 39×39×16 | 2×2 | (1,1) | 0 |
| MP1 | 39×39×16 | 19×19×16 | 2×2 | (2,2) | 0 |
| Conv2 | 19×19×16 | 18×18×32 | 2×2 | (1,1) | 0 |
| MP2 | 18×18×32 | 9×9×32 | 2×2 | (2,2) | 0 |
| FC1 | 2592 | 1024 | 1024×2592 | - | - |
| FC2 | 1024 | 1024 | 1024×1024 | - | - |
| FC3 | 1024 | 1964 | 1964×1024 | - | - |

| Vector2Image layer | | |
|---|---|---|
| Input dim. | Patch dim. | Number of patches in the background |
| 1964 | 4×4 | 364 |

Secondly, a CNN is constructed to capture the mapping relationship from $\{C_{v_1}^{reshape}, C_{v_2}^{reshape}\}$ to the unknown hierarchical temperature vector $\hat{T} = \begin{bmatrix} \hat{T}^{RoI} \\ \hat{T}^{BG} \end{bmatrix}$. Here, $\hat{T}^{RoI} \in \mathbb{R}^{J^{RoI} \times 1}$, $\hat{T}^{BG} \in \mathbb{R}^{(J-J^{RoI}) \times 1}$. As shown in Fig. 5, this CNN contains two convolutional layers, i.e. Conv1 and Conv2, two max-pooling layers, i.e. MP1 and MP2, and three fully-connected layers, i.e. FC1, FC2, and FC3. Table I shows the empirically determined hyper-parameters for the PI-CNN implemented in this work. The mathematical operations performed in these layers can be expressed as follows.

1) In the convolutional layers, the convolution operation is formulated by

$$O = g(W * I + b), \quad (9)$$

where $I \in \mathbb{R}^{H_I \times W_I \times C_I}$ is the input or intermediate feature maps, $O \in \mathbb{R}^{H_O \times W_O \times C_O}$ the output feature maps, $W \in \mathbb{R}^{H_W \times W_W \times C_W}$ the convolution kernel that performs as the weight matrix, $b \in \mathbb{R}^{C_b \times 1}$ the bias vector, * the operand for two-dimensional convolution, and $g(\cdot)$ the activation function. $H_I$ ($H_O$), $W_I$ ($W_O$), and $C_I$ ($C_O$) are the height, width, and channel of $I$ ($O$), respectively. $H_W$, $W_W$, and $C_W$ are the filter height, filter width, and the number of filters of $W$, respectively. $C_b$ is the length of $b$. In particular, $I \in \mathbb{R}^{H_{RoI} \times W_{RoI} \times 2}$ for Conv1 is formed by concatenating $C_{v_1}^{reshape}$ and $C_{v_2}^{reshape}$ into 3D matrix as

$$I = [C_{v_1}^{reshape}, C_{v_2}^{reshape}]. \quad (10)$$

2) In the max-pooling layers, the down-sampling output $o \in \mathbb{R}$ is given by

$$o = \max(I_{\text{filter}}), \quad (11)$$

where $I_{\text{filter}}$ denotes the filter with dimensions of $H_P \times W_P$.

3) In the fully-connected layers, the forward propagation is formulated by

$$O = g(WI + b), \quad (12)$$

where $I \in \mathbb{R}^{C_I \times 1}$, $O \in \mathbb{R}^{C_O \times 1}$, $W \in \mathbb{R}^{C_O \times C_I}$, $b \in \mathbb{R}^{C_I \times 1}$ and $g(\cdot)$ are the input vector, output vector, weight matrix, bias vector and activation function, respectively. In this network, ReLU is chosen as the activation function $g(\cdot)$.

Thirdly, the output from the CNN, i.e. $\hat{T}$, is transformed into the reconstructed hierarchical temperature image with Vetor2Image layer, according to Algorithm 2. The pixels in the RoI and the background are rearranged from $\hat{T}^{RoI}$ and $\hat{T}^{BG}$, respectively. Pixels out of the edges of the octagonal sensing region are set as zero.

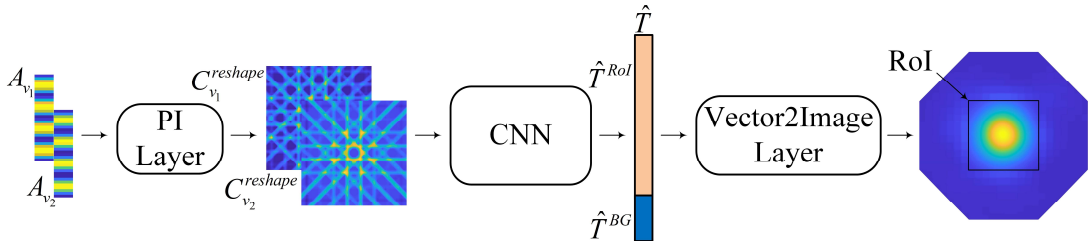

Fig. 4. The overall architecture of PI-CNN proposed in this work.

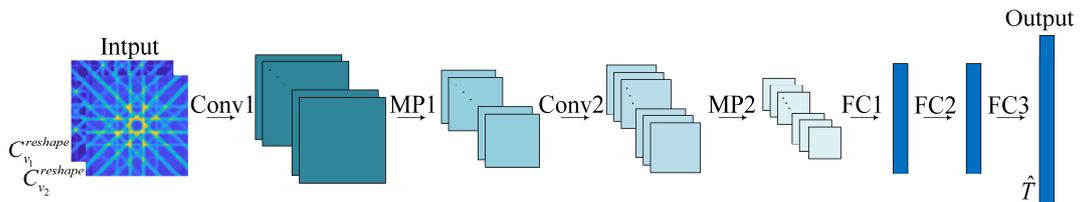

Fig. 5. The constructed CNN in the proposed PI-CNN.

**Algorithm 2** Image reconstruction from the hierarchical temperature vector $\hat{T}$ in the Vector2image layer

**Input**: Dimensions of the reconstructed temperature image $H_F \times W_F$, dimensions of the RoI $H_{RoI} \times W_{RoI}$, number of pixels in the RoI $J^{RoI}$, dimensions of patches in the background $H_{Patch} \times W_{Patch}$, number of patches in the background $J-J_{RoI}$, and hierarchical temperature vector $\hat{T} = \begin{bmatrix} \hat{T}^{RoI} \\ \hat{T}^{BG} \end{bmatrix}$.

**Initialise**: $\hat{F} \leftarrow$ zeros($H_F, W_F$)
**Operations**:
1: Reconstruct $\hat{F}$ in the RoI:
  $H_{RoI} \times W_{RoI}$ submatrix in $\hat{F}$ corresponding to the RoI $\leftarrow$ reshape($\hat{T}^{RoI}$, ($H_{RoI}$, $W_{RoI}$))
2: Reconstruct $\hat{F}$ in the background:
  **for** $j=1,\ldots,J-J^{RoI}$
    $j$-th patch in the background $\leftarrow$ duplicate($\hat{T}_j^{BG}$, ($H_{Patch}$, $W_{Patch}$))
  **end for**
**Output**: Reconstruction temperature image $\hat{F} \in \mathbb{R}^{H_F \times W_F}$.

## IV. NETWORK TRAINING AND TESTING

### A. Dataset Construction

Phantoms with the different combinations of 2D Gaussian inhomogeneities are constructed to simulate the multimodal flames encountered in practical combustion processes. Since $H_2O$ concentration is generally well-correlated with temperature in hydrocarbon flame [22], the distributions of temperature and $H_2O$ concentration, i.e. $T(x,y)$ and $X(x,y)$, are similarly modeled as

$$T(x,y) = \sum_{l=1}^{N_G} \lambda_l \cdot \frac{u_l}{\max[f_l(x,y)]} \cdot f_l(x,y) + T_{min} \quad (13)$$

and

$$X(x,y) = \sum_{l=1}^{N_G} \lambda_l \cdot \frac{v_l}{\max[f_l(x,y)]} \cdot f_l(x,y) + X_{min}, \quad (14)$$

where $(x,y)$ denotes the coordinates of the sensing region. $x \in \{1, 2,\ldots,H_F\}$ and $y \in \{1, 2,\ldots,W_F\}$. $l$ and $N_G$ denote the $l$-th and the total number of Gaussian inhomogeneity in the phantom, respectively. $T_{min}$ and $X_{min}$ are the minimum temperature and $H_2O$ concentration in the sensing region, respectively. $u_l$, $v_l$ and $\lambda_l$ are random scaling factors for the $l$-th Gaussian inhomogeneity. $u_l \sim U(T_{peakL}-T_{min}, T_{peakU}-T_{min})$, $v_l \sim U(X_{peakL}-X_{min}, X_{peakU}-X_{min})$, and $\lambda_l \sim U(0.7,1)$. Here, $U(a,b)$ denotes uniform distribution in range $(a,b)$. $T_{peakL}$ ($X_{peakL}$) and $T_{peakU}$ ($X_{peakU}$) are the lower and upper bounds of the peak values of temperature ($H_2O$ concentration) for each inhomogeneity, respectively. In other words, $T_{peakL}$ and $T_{peakU}$ set to tune randomly the peak of $u_l \cdot f_l(x,y)$ in the range of ($T_{peakL}, T_{peakU}$), while $X_{peakL}$ and $X_{peakU}$ set to tune randomly the peak of $v_l \cdot f_l(x,y)$ in the range of ($X_{peakL}, X_{peakU}$). $f_l(x,y)$ is the joint probability density function of the $l$-th Gaussian inhomogeneity expressed by

$$f_l(x,y) = \frac{1}{2\pi \sigma_x^l \sigma_y^l} \exp\left[-\left(\frac{(x-x_c^l)^2}{2(\sigma_x^l)^2} + \frac{(y-y_c^l)^2}{2(\sigma_y^l)^2}\right)\right],$$
$$l = 1,\ldots,N_G \quad (15)$$

where $(x_c^l, y_c^l)$ is the central position of the $l$-th Gaussian inhomogeneity. $\sigma_x^l$ and $\sigma_y^l$ are the standard deviations along $x$ and $y$ axes of the $l$-th Gaussian inhomogeneity, respectively.

As illustrated previously in Section III, the sensitivity matrix $L$ is determined by the optical layout of the 32-beam TDLAS tomographic senor. Given the phantoms of 2D temperature and $H_2O$ concentration, path integrated absorbances $A_{v_1} \in \mathbb{R}^{32 \times 1}$ and $A_{v_2} \in \mathbb{R}^{32 \times 1}$ can be calculated according to the forward formulation described by (2). Finally, a sample, noted as (($A_{v_1}$, $A_{v_2}$), $T$), is generated as a combination of the hierarchical temperature distribution $T$ and the measurements without noise contamination on ($A_{v_1}, A_{v_2}$).

In this work, a dataset with a total of 10,900 samples is artificially created, including 4,500 single-peak flames ($N_G$=1) and 6,400 double-peak flames ($N_G$=2). It is then randomly divided into a training set with 10,000 samples and a test set with 900 samples. Parameters of samples are set as: $T_{min}$=300 K, $T_{peakL}$=600 K, $T_{peakU}$=900 K, $X_{min}$=0.01, $X_{peakL}$=0.10, and $X_{peakU}$=0.12. To cover the majority of the Gaussian inhomogeneities within the sensing field, $x_c^l$ and $y_c^l$ are randomly selected from uniform distribution $U(34, 65)$, while $\sigma_x^l$ and $\sigma_y^l$ are randomly selected from $U(10, 25)$.

### B. Network Training

To examine the performance of the proposed PI-CNN, it is compared to the following two state-of-the-art CNNs with similar architectures:

1) Huang's CNN (H-CNN). As shown in Fig. 6, the CNN proposed by Huang et al. in [14], named as H-CNN, contains two convolutional layers, i.e. Conv1 and Conv2, one average pooling layer, i.e. AP, and one fully-connected layer, i.e. FC. The leaky ReLU activation function is adopted [23]. It originally takes LoS measurements as the inputs, and reconstructs the temperature distribution with uniform-size pixels in the RoI. Here, H-CNN is adapted for hierarchical imaging of temperature distribution in the whole sensing region by employing $A_{v_1}^{reshape} \in \mathbb{R}^{4 \times 8}$ and $A_{v_2}^{reshape} \in \mathbb{R}^{4 \times 8}$ as the inputs, and the hierarchical temperature distribution $\hat{T}$ as the output. The main architecture of H-CNN is kept in the flow diagram with the above-mentioned signal preprocessing added to adapt the inputs for this specific work. Experimentally determined hyper-parameters of this network are shown in Table II.

2) Direct CNN (D-CNN). To address the effectiveness of the proposed pseudo-inversion, the CNN (without the PI layer), named as D-CNN, is extracted from PI-CNN shown in Fig. 4 and used as an individual network in the comparative study. As shown in Fig. 7, the hierarchical temperature image is directly reconstructed from $A_{v_1}^{reshape} \in \mathbb{R}^{4 \times 8}$ and $A_{v_2}^{reshape} \in \mathbb{R}^{4 \times 8}$. Due to the small dimension of feature maps output from convolutional layers Conv1 and Conv2, pooling layers are removed to preserve spatial information. Other hidden layers are the same as those in PI-CNN. Table III shows the empirically determined hyper-parameters of the D-CNN.

All these three networks, i.e. H-CNN, D-CNN and the proposed PI-CNN, are trained on the same training set with 10,000 noise-free samples constructed in Section IV. A. The $L_2$ loss function is defined as

$$L_2 = \frac{1}{B}\sum_{b=1}^{B}\left\|\hat{\pmb{T}}_b - \pmb{T}_b\right\|_2 \quad (16)$$

where $B$ is the batch size, set as 128 in this work, $\hat{\pmb{T}}_b$ and $\pmb{T}_b$ are the reconstructed and true hierarchical temperature vectors of the $b$-th training sample in the batch, respectively. Adam optimizer [24] is employed with an empirically determined learning rate of 0.001. $L_2$ regularization is employed to prevent the networks from overfitting with the penalty factor set as 0.0001. The number of epochs is set as 100 for adequate convergence. All the three networks are implemented with Deep Learning Toolbox (14.0) in Matlab (R2020a) on a PC with an Intel Core i7-9700 3.00 GHz CPU and 16 GB memory.

### C. Test Results

To examine the performances of the proposed neural networks with practical measurements, additional Gaussian noise is added to $A_v$ in the test set as follows:

$$\pmb{A}_v += \sigma \cdot \pmb{n} \quad (17)$$

where $\pmb{n} \in \mathbb{R}^{32\times1}$ is the noise vector conforming to standard distribution $G(0,1)$, and $\sigma$ the standard deviation of the Gaussian noise converted from the specified Signal to Noise Ratio (SNR). Seven test sets with different levels of noise are generated by ranging SNR from 20 dB to 50 dB with a step of 5 dB. The hierarchical temperature images are reconstructed by feeding the noise-contaminated ($A_{v_1}, A_{v_2}$) of the test samples into the optimally trained H-CNN, D-CNN, and PI-CNN, respectively.

Firstly, the hierarchical temperature images reconstructed by H-CNN, D-CNN and PI-CNN are visually inspected under a practical SNR of 35 dB in real applications. Fig. 8 (a) and Fig. 9 (a) show two representative phantoms with a single Gaussian inhomogeneity and two adjacent Gaussian inhomogeneities, respectively. Figs. 8 (b-d) and Figs. 9 (b-d) shows their reconstructions using H-CNN, D-CNN and PI-CNN, respectively. For the single-inhomogeneity case, all the three algorithms can reliably indicate the profile of the inhomogeneity with correct location of the peak in the sensing region. However, the PI-CNN gives the best fidelity with very few artefacts in the reconstructed image. For the double-inhomogeneity case, the PI-CNN shows much stronger superiority than the other two algorithms. The temperature image reconstructed using H-CNN (Fig. 9 (b)) underestimates the widths of both inhomogeneities and obvious distortion of the temperature profile, while that reconstructed using D-CNN (Fig. 9 (c)) overestimates the width of the right-hand inhomogeneity.

Then, the performances of the three algorithms are quantitatively evaluated under different SNRs by introducing three metrics, Relative Peak Distance Error (RPDE), Relative Peak Amplitude Error (RPAE), and Relative Reconstruction Error (RRE). To be specific, RPDE and RPAE characterize the dislocation and offset of the reconstructed peaks of the inhomogeneities from the true ones, respectively.

RPDE is defined as

$$d^{peak} = \frac{1}{N_G}\sum_{l=1}^{N_G}\frac{\sqrt{\left(x_r^l - x_c^l\right)^2 + \left(y_r^l - y_c^l\right)^2}}{R}, \quad (18)$$

TABLE II
HYPER-PARAMETERS OF THE H-CNN IMPLEMENTED IN THIS WORK.

|       | Input dim. | Output dim. | weight matrix size | Stride | Padding |
|-------|------------|-------------|--------------------|--------|---------|
| Conv1 | 8×4×2      | 7×3×8       | 2×2                | (1,1)  | 0       |
| AP    | 7×3×8      | 6×2×8       | 2×2                | (1,1)  | 0       |
| Conv2 | 6×2×8      | 5×1×14      | 2×2                | (1,1)  | 0       |
| FC    | 70         | 1964        | 1964×70            | -      | -       |

TABLE III
HYPER-PARAMETERS OF THE D-CNN IMPLEMENTED IN THIS WORK.

|       | Input dim. | Output dim. | weight matrix size | Stride | Padding |
|-------|------------|-------------|--------------------|--------|---------|
| Conv1 | 8×4×2      | 7×3×16      | 2×2                | (1,1)  | 0       |
| Conv2 | 7×3×16     | 6×2×32      | 2×2                | (1,1)  | 0       |
| FC1   | 384        | 1024        | 1024×384           | -      | -       |
| FC2   | 1024       | 1024        | 1024×1024          | -      | -       |
| FC3   | 1024       | 1964        | 1964×1024          | -      | -       |

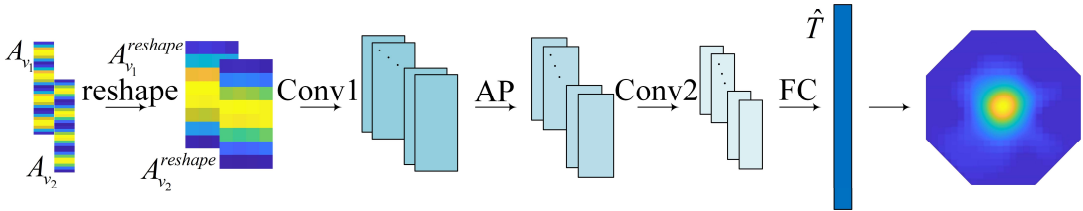

Fig. 6. The architecture of the H-CNN adapted for hierarchical temperature imaging.

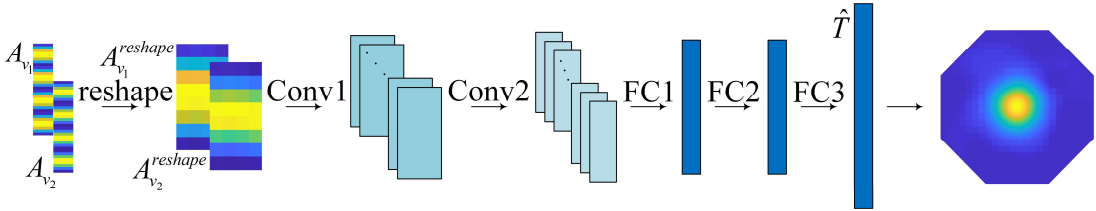

Fig. 7. The architecture of D-CNN.

where $(x_r^l, y_r^l)$ is the coordinate of the reconstructed peak of the $l$-th Gaussian inhomogeneity. $R$ is the side length of the octagonal sensing region.

RPAE is defined as

$$e^{peak} = \frac{1}{N_G} \sum_{l=1}^{N_G} \frac{\left|T(x_r^l, y_r^l) - T(x_c^l, y_c^l)\right|}{\left|T(x_c^l, y_c^l)\right|}, \quad (19)$$

where $T(x_c^l, y_c^l)$ and $T(x_r^l, y_r^l)$ are the amplitudes of the true and reconstructed peaks of the $l$-th Gaussian inhomogeneity, respectively.

RRE performs as an overall evaluation of the image quality by quantifying the pixel-wised reconstruction error. RRE between the reconstructed hierarchical temperature image $\hat{T}$ and the ground truth $T$ is defined as

$$e^T = \frac{\|\hat{T} - T\|_2}{\|T\|_2}. \quad (20)$$

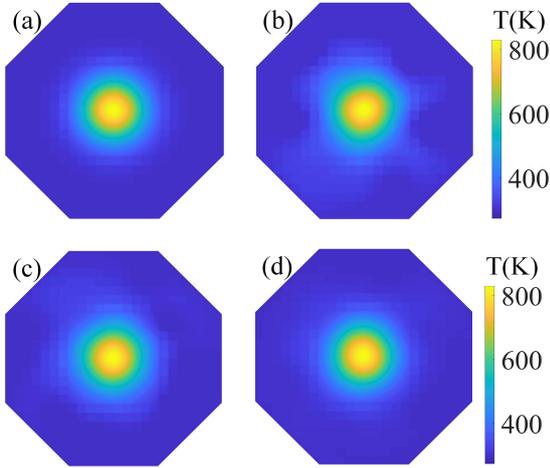

Fig. 8. Reconstruction of hierarchical temperature images for a single-inhomogeneity phantom under an SNR of 35 dB. (a) shows the true temperature image. (b), (c) and (d) shows the hierarchical temperature images reconstructed using H-CNN, D-CNN and PI-CNN, respectively.

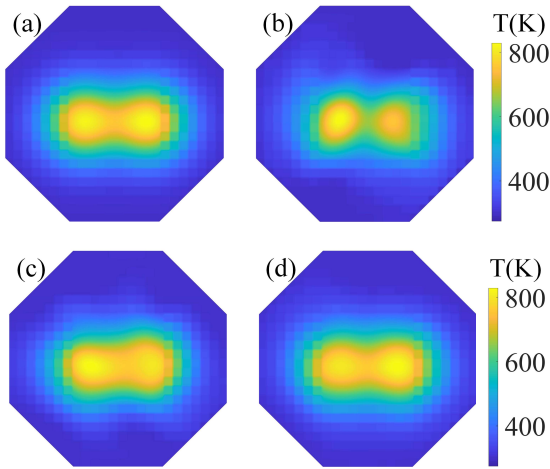

Fig. 9. Reconstruction of hierarchical temperature images for a double-inhomogeneity phantom under an SNR of 35 dB. (a) shows the true temperature image. (b), (c) and (d) shows the hierarchical temperature images reconstructed using H-CNN, D-CNN and PI-CNN, respectively.

As illustrated in Section IV.A, there are 900 samples in each test set. Therefore, RPDE, RPAE and RRE can be obtained for each sample. To statistically characterize the performance of the algorithms for the test set, RPDEs, RPAEs and RREs over the 900 samples are averaged by

$$D^{peak} = \frac{1}{H} \sum_{h=1}^{H} d_h^{peak}, \quad (21)$$

$$E^{peak} = \frac{1}{H} \sum_{h=1}^{H} e_h^{peak}, \quad (22)$$

$$E^T = \frac{1}{H} \sum_{h=1}^{H} e_h^T, \quad (23)$$

where $d_h^{peak}$, $e_h^{peak}$ and $e_h^T$ are the RPDE, RPAE and RRE of the $h$-th sample in the test set, respectively. $H$ is the total number of samples in the test set. Here, $H = 900$.

Figs. 10 and 11 depict $D^{peak}$ and $E^{peak}$ obtained using the three networks for the test set at different levels of noise, respectively. For all the three networks, it can be seen both $D^{peak}$ and $E^{peak}$ decrease as SNR increases. At all levels of noise, PI-CNN always achieves the lowest $D^{peak}$ and $E^{peak}$, indicating its better agreement with the ground truth in terms of both the peak locations and amplitudes of the Gaussian inhomogeneities in the reconstructed images. In addition, D-CNN performs better than H-CNN under different SNRs. $D^{peak}$ and $E^{peak}$ obtained using D-CNN are half of those obtained using H-CNN when the SNR is greater than 35 dB.

Fig. 12 shows the dependence of $E^T$ on SNRs for the three algorithms. Similar as $D^{peak}$ vs. SNRs and $E^{peak}$ vs. SNRs, $E^T$ decreases as SNR increases for all the three networks. Under very noisy conditions, for example, SNR = 20 dB, similar $E^T$ are obtained using H-CNN and D-CNN. As SNR increases, $E^T$ obtained using PI-CNN is always the lowest, while that obtained using H-CNN is the highest. Taking the SNR of 35 dB as an example, $E^T$ obtained using H-CNN, D-CNN and PI-CNN are 0.060, 0.033 and 0.022, respectively. That is, in comparison with H-CNN and D-CNN, PI-CNN suppresses the reconstruction errors by 63.33% and 33.33%, respectively. It can be concluded the proposed PI-CNN is more capable of retrieving high-fidelity hierarchical temperature images.

Finally, the time elapsed for training and single-frame image reconstruction using H-CNN, D-CNN and PI-CNN are compared in Table IV. The training time of PI-CNN is 26.1 minutes, which is around 3 times longer than that of D-CNN and 11 times longer than that of H-CNN. This is caused by the larger dimensions of features maps fed into layers of PI-CNN. In terms of reconstruction time, PI-CNN performs similar with the other two networks by completing a single reconstruction within 0.5 millisecond. It is worth mentioning the training is a one-off process, while the image reconstruction is a repetitive process. In practical applications of TDLAS tomography, there are generally thousands or even millions of time-dependent temperature distributions to be repetitively reconstructed to characterize the dynamics of the target flows and combustion processes. Therefore, the importance of reconstruction time significantly outweighs that of the one-off training time. Regarding the PI-CNN proposed in this paper, more than 2,000 frames of images can be reconstructed within 1 second, indicating strong real-time performance of the proposed network.

TABLE IV
COMPARISON OF THE TIME ELAPSED FOR TRAINING AND SINGLE-FRAME IMAGE RECONSTRUCTION USING H-CNN, D-CNN AND PI-CNN.

| Network | Training time (min) | Reconstruction time of a test sample (sec) |
|---|---|---|
| H-CNN | 2.3 | 3.949e-04 |
| D-CNN | 8.7 | 3.569e-04 |
| PI-CNN | 26.1 | 4.854e-04 |

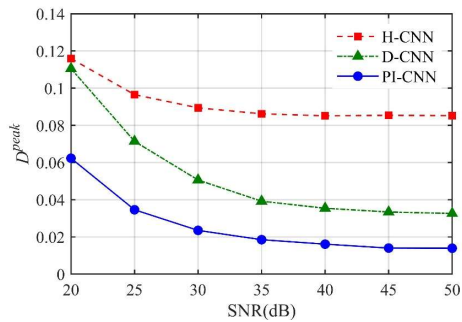

Fig. 10. Dependence of $D^{peak}$ on SNRs for H-CNN, D-CNN and PI-CNN.

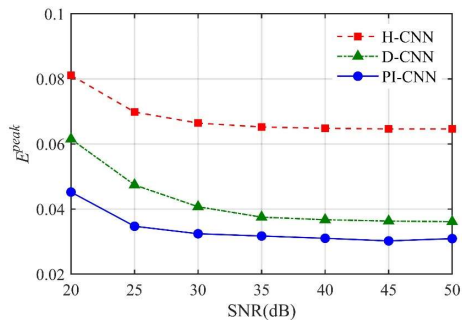

Fig. 11. Dependence of $E^{peak}$ on SNRs for H-CNN, D-CNN and PI-CNN.

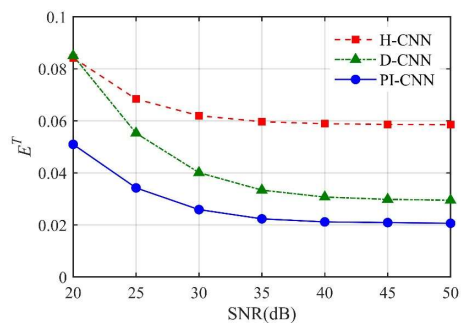

Fig. 12. Dependence of $E^T$ on SNRs for H-CNN, D-CNN, and PI-CNN.

## V. EXPERIMENT

Lab-scale experiments were carried out to further validate the proposed PI-CNN. The 32-beam TDLAS tomographic sensor was built with the optical layout shown in Fig. 2. More details of the optics and hardware electronics have been described in our previous publications [7, 21]. In addition, Wavelength Modulation Spectroscopy (WMS) was used for each of the LoS-TDLAS measurement, the parameter settings in WMS have been detailed in [7].

Similar as the simulation, cases with single-inhomogeneity and double-inhomogeneity temperature distributions are demonstrated in the experiment. Fig. 13 (a) shows the single-inhomogeneity case with a flame located at the center of the sensing region, while Fig. 14 (a) shows the same flame relocated at the left center of the sensing region. Fig. 15 (a) shows the double-inhomogeneity case with a larger-size flame located lower left and a smaller-size flame upper right. For all the three cases, the peaks of inhomogeneities are located in the RoI with dense pixels.

For the single-inhomogeneity case of Fig. 13 (a), the hierarchical temperature images reconstructed by H-CNN, D-CNN and PI-CNN are shown in Figs. 13 (b), (c), and (d), respectively. For each of the three reconstructed images, the retrieved peak position agrees well with that in the phantom. The amplitudes of the peak temperature are also consistent in the three images, giving 669 K, 804 K and 753 K in Figs. 13 (b), (c) and (d), respectively. However, the temperature image retrieved using the proposed PI-CNN has fewer artefacts in comparison with those retrieved using H-CNN and D-CNN. The blurred boundary of the reconstructed inhomogeneity in Fig. 13 (d) is mainly caused by the heat dissipation from the central hot zone.

For the single-inhomogeneity case of Fig. 14 (a), the hierarchical temperature images reconstructed by H-CNN, D-CNN and PI-CNN are shown in Figs. 14 (b), (c), and (d), respectively. The relocated flame can be clearly localized using H-CNN, D-CNN and PI-CNN, respectively. The amplitudes of the peak temperature are 543 K, 760 K, and 751 K in Figs. 14 (b), (c) and (d), with offsets of 126 K, 44 K, and 2 K, compared with those in Figs. 13 (b), (c) and (d), respectively. The retrieved peak temperature using the proposed PI-CNN shows best agreement with the case in Fig. 13. In addition, the temperature image reconstructed by PI-CNN has fewer artefacts compared with that reconstructed using D-CNN.

For the double-inhomogeneity case, the reconstruction of the larger-size flame is convincing for all the three algorithms, showing correct locations and peak amplitudes of 774 K, 779 K and 770 K in Figs. 15 (b), (c) and (d), respectively. However, the retrieval of the smaller-size flame suffer from the interference by its neighboring artifacts when using H-CNN and D-CNN. In Figs. 15 (b), (c) and (d), the amplitudes of the peak temperature for the smaller-size flame are 532 K, 611 K and 752 K, respectively. As the smaller-size flame is the same as that used in Fig. 13 and Fig. 14, the retrieved peak temperature using the PI-CNN shows best agreement with those single-inhomogeneity cases.

## VI. CONCLUSION

A novel hierarchical imaging strategy using PI-CNN aided TDLAS tomography is proposed in this paper. This new development not only prioritizes the training and learning resources for temperature imaging of the target flow in the RoI with high spatial resolution, but also maintains the integrity of the spectroscopic absorption model by reconstructing the less spatially resolved background temperature. The introduced pseudo inversion reveals inherent correlation between TDLAS measurements and the hierarchical temperature distribution, thus facilitating better-fidelity image reconstruction.

The performance of PI-CNN is firstly evaluated using the simulated test sets, and compared with two existing networks, i.e. H-CNN and D-CNN. Simulation results indicate the PI-CNN has strong robustness and outperforms the other two networks for a wide range of SNR. In the lab-scale experiments, the single-inhomogeneity and double-inhomogeneity temperature distributions reconstructed using the proposed PI-CNN achieve not only the strongest noise resistance, but also the best agreement with the original phantoms in terms of both the position and amplitude of the peak temperature. In addition, PI-CNN is capable of reconstructing more than 2,000 frames of images per second, indicating great potential for high-speed and real-time temperature imaging in industrial applications of TDLAS tomography.

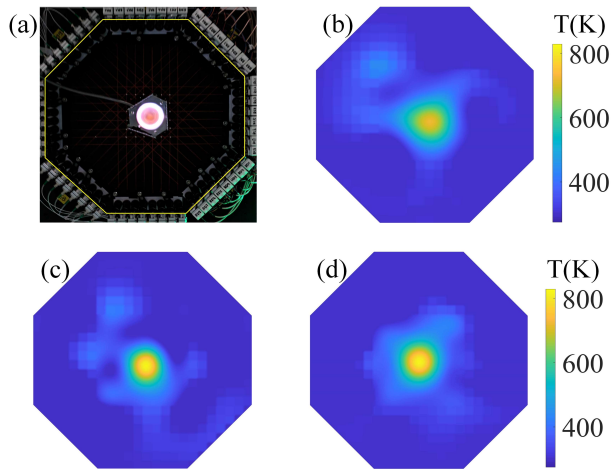

Fig. 13. Reactive flow field generated in the experiment with (a) a single inhomogeneity at the center and the hierarchical temperature image reconstructed using (b) H-CNN, (c) D-CNN and (d) PI-CNN, respectively.

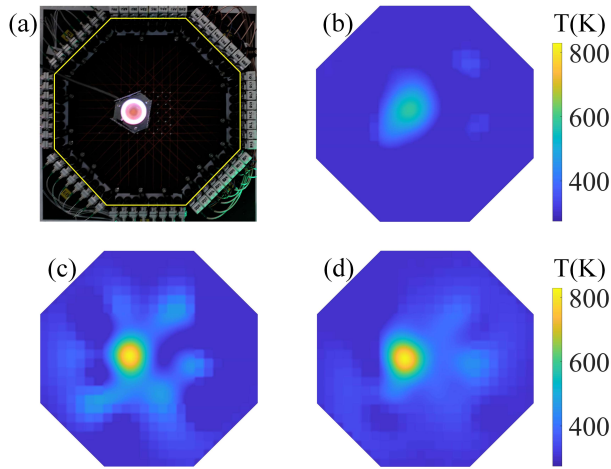

Fig. 14. Reactive flow field generated in the experiment with (a) a single inhomogeneity relocated at the left center and the hierarchical temperature image reconstructed using (b) H-CNN, (c) D-CNN and (d) PI-CNN, respectively.

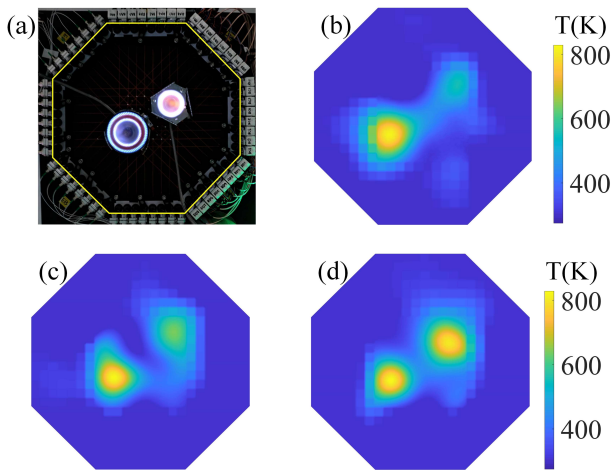

Fig. 15. Reactive flow field generated in the experiment with (a) two inhomogeneities and the hierarchical temperature images reconstructed using (b) H-CNN, (c) D-CNN and (d) PI-CNN, respectively.